%% file: ISITA2020_Polarsubcodes_v15_ArxivversionLong.tex
\documentclass[conference,10pt,letterpaper]{IEEEtran}

\usepackage[utf8]{inputenc} 
\usepackage[T1]{fontenc}
\usepackage{cite}
\usepackage[cmex10]{amsmath} 

\usepackage{booktabs}

\DeclareFontFamily{U}{mathx}{\hyphenchar\font45}
\DeclareFontShape{U}{mathx}{m}{n}{<-> mathx10}{}
\DeclareSymbolFont{mathx}{U}{mathx}{m}{n}
\DeclareMathAccent{\widebar}{0}{mathx}{"73}

\usepackage{amsmath,amsfonts,amssymb,psfrag}
\usepackage{amsthm}
\usepackage{graphicx}
\usepackage{epstopdf}
\usepackage{rotating}
\usepackage{bm}
\usepackage{dsfont}
\usepackage{color}
\usepackage{tikz}
\usetikzlibrary{plotmarks}
\usetikzlibrary{shapes,arrows,fit,calc,positioning,automata}
\usepackage{pgfplots}
\usetikzlibrary{calc}
\usetikzlibrary{decorations.markings}
\usetikzlibrary{positioning}
\pgfplotsset{compat=1.10}

\makeatletter
\newsavebox\myboxA
\newsavebox\myboxB
\newlength\mylenA

\newcommand*\xoverline[2][0.75]{%
	\sbox{\myboxA}{$\m@th#2$}%
	\setbox\myboxB\null
	\ht\myboxB=\ht\myboxA%
	\dp\myboxB=\dp\myboxA%
	\wd\myboxB=#1\wd\myboxA
	\sbox\myboxB{$\m@th\overline{\copy\myboxB}$}
	\setlength\mylenA{\the\wd\myboxA}
	\addtolength\mylenA{-\the\wd\myboxB}%
	\ifdim\wd\myboxB<\wd\myboxA%
	\rlap{\hskip 0.5\mylenA\usebox\myboxB}{\usebox\myboxA}%
	\else
	\hskip -0.5\mylenA\rlap{\usebox\myboxA}{\hskip 0.5\mylenA\usebox\myboxB}%
	\fi}
\makeatother

\usepackage{mathtools}
\usepackage{verbatim}
\usepackage[ruled,vlined,titlenumbered,linesnumbered]{algorithm2e}
\usepackage{tabularx}
\usepackage{makecell}
\usepackage{multirow}
\usepackage{dsfont}
\usepackage{array,bm}
\newcolumntype{L}[1]{>{\raggedright\let\newline\\\arraybackslash\hspace{0pt}}m{#1}}
\newcolumntype{C}[1]{>{\centering\let\newline\\\arraybackslash\hspace{0pt}}m{#1}}
\newcolumntype{R}[1]{>{\raggedleft\let\newline\\\arraybackslash\hspace{0pt}}m{#1}}

\newtheorem{remark}{Remark}
\newtheorem{definition}{Definition}

\newtheorem{corollary}{Corollary}

\newcommand{\pB}{\ensuremath{P_{\text{B}}}}

\newcommand{\Enc}{\mathsf{Enc}}
\newcommand{\Dec}{\mathsf{Dec}}

\newcommand*\xor{\mathbin{\oplus}}

\begin{document}
\title{Randomized Nested Polar Subcode Constructions for Privacy, Secrecy, and Storage}

 \author{%
   \IEEEauthorblockN{Onur G\"unl\"u\textsuperscript{1},
                     Peter Trifonov\textsuperscript{2},
                     Muah Kim\textsuperscript{1},
                     Rafael F. Schaefer\textsuperscript{1},
                     and Vladimir Sidorenko\textsuperscript{3}}
   \IEEEauthorblockA{\textsuperscript{1}%
                     Information Theory and Applications Chair,                     TU Berlin, Germany,
                     \{guenlue, muah.kim, rafael.schaefer\}@tu-berlin.de}
   \IEEEauthorblockA{\textsuperscript{2}%
                     Faculty of Secure Information Technologies, ITMO, Russia,
                     pvtrifonov@itmo.ru}
   \IEEEauthorblockA{\textsuperscript{3}%
                     Chair of Communications Engineering, TU Munich, Germany,
                     vladimir.sidorenko@tum.de}
 }

\maketitle

\begin{abstract}
	We consider polar subcodes (PSCs), which are polar codes (PCs) with dynamically-frozen symbols, to increase the minimum distance as compared to corresponding PCs. A randomized nested PSC construction with a low-rate PSC and a high-rate PC, is proposed for list and sequential successive cancellation decoders. This code construction aims to perform lossy compression with side information. Nested PSCs are used in the key agreement problem with physical identifiers. Gains in terms of the secret-key vs. storage rate ratio as compared to nested PCs with the same list size are illustrated to show that nested PSCs significantly improve on nested PCs. The performance of the nested PSCs is shown to improve with larger list sizes, which is not the case for nested PCs considered.
\end{abstract}

\IEEEpeerreviewmaketitle
\section{Introduction}
A common secrecy problem considers the wiretap channel (WTC) \cite{WynerWTC}. The WTC encoder aims to hide a transmitted message from an eavesdropper with a channel output correlated with the observation of a legitimate receiver. There are various code constructions for the WTC that achieve the secrecy capacity, e.g., in \cite{WTCpolarVardy,KliewerWTC,OzanWTC,YingbingWTC}. Some of these constructions use nested polar codes (PCs) \cite{Arikan}, which have a low encoding/decoding complexity, asymptotic optimality for various problems, and good finite length performance if a successive cancellation list (SCL) decoder in combination with an outer cyclic redundancy check (CRC) code are used \cite{SCLPolar}. Similarly, nested PCs achieve the strong coordination capacity boundaries \cite{strongcoordinationPolar}; see, e.g., \cite{RemiStrongCoordination}. 

A closely related secrecy problem to the WTC problem is the key agreement problem with two terminals that observe correlated random variables and have access to a public, authenticated, and one-way communication link; whereas an eavesdropper observes only the public messages called \textit{helper data} \cite{AhlswedeCsiz, Maurer}. There are two common models for key agreement: the \textit{generated-secret (GS)} model, where an encoder extracts a secret key from the sequence observed, and the \textit{chosen-secret (CS)} model, where a pre-determined secret key is given as input to the encoder. The main constraint for this problem is that the construction should not leak information about the secret key (negligible \textit{secrecy leakage}). Furthermore, a \textit{privacy leakage} constraint is introduced in \cite{IgnaTrans} to leak as little information about the identifier as possible. Similarly, \textit{storage} in the public communication link can be expensive and limited, e.g., for internet-of-things (IoT) device applications \cite{csiszarnarayan,bizimWZ}. The regions of achievable secret-key vs. privacy-leakage (key-leakage) rates for the GS and CS models are given in \cite{IgnaTrans}, while the key-leakage-storage regions with multiple encoder measurements are treated in \cite{bizimMMMMTIFS}.

An important application of these key agreement models is the key agreement with physical identifiers such as digital circuits that have outputs unique to the device that embodies them. Examples of these physical identifiers are physical unclonable functions (PUFs) \cite{GassendThesis, PappuThesis}. The start-up behavior of static random access memories (SRAM) and the speckle pattern observed from coherent waves propagating through a disordered medium can serve as PUFs that have reliable outputs and high entropy \cite{IgnaCTW, benimdissertation}. 

Optimal nested random linear code constructions for the lossy source coding with side information problem, i.e., Wyner-Ziv (WZ) problem \cite{WZ}, are shown in \cite{bizimWZ} to be optimal also for the key agreement with PUFs. Thus, nested PCs are designed in \cite{bizimWZ} for practical SRAM PUF parameters to illustrate that nested PCs achieve rate tuples that cannot be achieved by using previous code constructions. The finite length performance of the nested PCs designed in \cite{bizimWZ} without an outer CRC code is not necessarily good due to small minimum distance of PCs. Therefore, we propose to increase the minimum distance by using PCs with dynamically-frozen symbols (DFSs), i.e., \textit{polar subcodes (PSCs)} \cite{PolarSubcodesFirst}. 
 
PSCs represent a generalization of PCs, where frozen symbols are set to  linear combinations of other symbols. In general, randomized polar subcodes  \cite{PeterRandomizedPolarSubcode}  provide better performance than algebraic polar subcodes \cite{PolarSubcodesFirst} under list or sequential decoding with small list size. We therefore design codes for key agreement with PUFs by constructing nested PSCs in a randomized manner. Nested codes have a broad use, e.g., in WTC and strong coordination problems, so the proposed construction might be useful also for these problems. A summary of the main contributions is as follows.

We propose a method to obtain nested PSCs used as a WZ-coding construction. Furthermore, we develop a design procedure for the proposed construction adapted to the problem of key agreement with physical identifiers. Consider binary symmetric sources (BSSs) and channels (BSCs). Ring oscillator (RO) PUFs with transform coding \cite{bizimKittipongTIFS} and SRAM PUFs \cite{maes2009soft} are modeled by these sources and channels. We design and simulate nested PSCs for SRAM PUFs to illustrate that nested PSCs with sequential successive cancellation (SC) decoders for a list size of $L\!=\!8$ achieve significantly larger key vs. storage rate ratio than previously-proposed codes including nested PCs from \cite{bizimWZ} that approach the maximum likelihood (ML) performance with an SCL decoder for $L\!=\!8$. Nested PSC performance is illustrated to further improve with larger but reasonable list sizes such as $L\!=\!32,64$. 

This paper is organized as follows. In Section~\ref{sec:problem_settingandcode}, we describe the GS and CS models, and evaluate the key-leakage-storage region for BSSs and BSCs. We propose a randomized nested PSC construction and a design procedure adapted to key agreement with PUFs in Section~\ref{sec:nestedPSCforPUFs}. Significant key vs. storage rate ratio gains from nested PSCs as compared to previously-proposed codes are illustrated in Section~\ref{sec:SRAMPUFperformance}. 

\section{Problem Formulation}\label{sec:problem_settingandcode}
\begin{figure}
	\centering
	\resizebox{0.99\linewidth}{!}{
		\begin{tikzpicture}
		\node (so) at (-1.5,-2.2) [draw,rounded corners = 5pt, minimum width=1.0cm,minimum height=0.8cm, align=left] {$P_X(\cdot)$};
		\node (a) at (0,0) [draw,rounded corners = 6pt, minimum width=3.2cm,minimum height=1.2cm, align=left] {$
			(S,W) \overset{(a)}{=} \Enc\left(X^n\right)$\\ $W'\overset{(b)}{=}\Enc\left(X^n,S'\right)$};
		\node (c) at (3,-2.2) [draw,rounded corners = 5pt, minimum width=1.6cm,minimum height=0.8cm, align=left] {$P_{Y|X}(\cdot)$};
		\node (b) at (6,0) [draw,rounded corners = 6pt, minimum width=3.2cm,minimum height=1.2cm, align=left] {$\widehat{S} \overset{(a)}{=} \Dec\left(Y^n,W\right)$\\$\widehat{S}' \overset{(b)}{=} \Dec\left(Y^n,W'\right)$};
		\draw[decoration={markings,mark=at position 1 with {\arrow[scale=1.5]{latex}}},
		postaction={decorate}, thick, shorten >=1.4pt] (a.east) -- (b.west) node [midway, above] {$(a) W$} node [midway, below] {$(b) W'$};;
		\node (a1) [below of = a, node distance = 2.2cm] {$X^n$};
		\node (b1) [below of = b, node distance = 2.2cm] {$Y^n$};
		\node (k9) [below of = a1, node distance = 0.6cm] {Enrollment};
		\node (k19) [below of = b1, node distance = 0.6cm] {Reconstruction};
		\draw[decoration={markings,mark=at position 1 with {\arrow[scale=1.5]{latex}}},
		postaction={decorate}, thick, shorten >=1.4pt] (so.east) -- (a1.west);
		\draw[decoration={markings,mark=at position 1 with {\arrow[scale=1.5]{latex}}},
		postaction={decorate}, thick, shorten >=1.4pt] (a1.north) -- (a.south);
		\draw[decoration={markings,mark=at position 1 with {\arrow[scale=1.5]{latex}}},
		postaction={decorate}, thick, shorten >=1.4pt] (a1.east) -- (c.west);
		\draw[decoration={markings,mark=at position 1 with {\arrow[scale=1.5]{latex}}},
		postaction={decorate}, thick, shorten >=1.4pt] (c.east) -- (b1.west);
		\draw[decoration={markings,mark=at position 1 with {\arrow[scale=1.5]{latex}}},
		postaction={decorate}, thick, shorten >=1.4pt] (b1.north) -- (b.south);
		\node (a2) [above of = a, node distance = 2.2cm] {$S\quad\,\, S'$};
		\node (b2) [above of = b, node distance = 2.2cm] {$\widehat{S}\quad\,\,\widehat{S}'$};
		\draw[decoration={markings,mark=at position 1 with {\arrow[scale=1.5]{latex}}},
		postaction={decorate}, thick, shorten >=1.4pt] ($(b.north)-(0.3,0)$) -- ($(b2.south)-(0.3,0)$) node [midway, left] {$(a)$};
		\draw[decoration={markings,mark=at position 1 with {\arrow[scale=1.5]{latex}}},
		postaction={decorate}, thick, shorten >=1.4pt] ($(b.north)+(0.3,0)$)-- ($(b2.south)+(0.3,0)$) node [midway, right] {$(b)$};
		\draw[decoration={markings,mark=at position 1 with {\arrow[scale=1.5]{latex}}},
		postaction={decorate}, thick, shorten >=1.4pt] ($(a.north)-(0.3,0)$)-- ($(a2.south)-(0.3,0)$) node [midway, left] {$(a)$};
		\draw[decoration={markings,mark=at position 1 with {\arrow[scale=1.5]{latex}}},
		postaction={decorate}, thick, shorten >=1.4pt]  ($(a2.south)+(0.3,0)$)-- ($(a.north)+(0.3,0)$) node [midway, right] {$(b)$};
		\end{tikzpicture}
	}
	\caption{The $(a)$ GS and $(b)$ CS models.}\label{fig:problemsetup}
\end{figure}
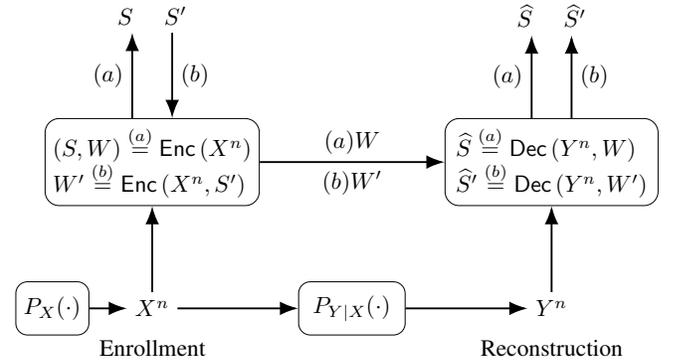

An identifier output is used to generate a secret key in the GS model, depicted in Fig.~\ref{fig:problemsetup}$(a)$. The source $\mathcal{X}$, noisy measurement $\mathcal{Y}$, secret key $\mathcal{S}$, and storage $\mathcal{W}$ alphabets are finite sets. During enrollment, the encoder observes the i.i.d. identifier output $X^n$ and computes a secret key $S$ and public helper data $W$ as $\displaystyle (S,W)\,{=}\,{\Enc}(X^n)$. During reconstruction, the decoder observes a noisy source measurement $Y^n$ of the source output $X^n$ through a memoryless measurement channel $P_{Y|X}$ in addition to the helper data $W$. The decoder estimates the secret key as $\displaystyle \widehat{S}\,{=}\,{\Dec}(Y^n\!,W)$. Fig.~\ref{fig:problemsetup}$(b)$ shows the CS model, where a secret key $S'\in\mathcal{S}$ is embedded into the helper data as $W' = \Enc(X^n,S')$. The decoder for the CS model estimates the secret key as $\widehat{S}'=\Dec(Y^n,W')$. Since the analyses for the CS model follows from the analyses for the GS model, it suffices to consider the GS model to illustrate the performance gains from nested PSCs.



\begin{definition}\label{def:achievabilityGSCS}
	A key-leakage-storage tuple $(R_s,R_\ell,R_w)$ is \emph{achievable} for the GS model if, given any $\epsilon>0$, there is some $n\!\geq\!1$, an encoder, and a decoder such that $R_s=\frac{\log|\mathcal{S}|}{n}$ and
	\begin{align}
	&\pB\;\triangleq\;\Pr[\widehat{S} \neq S] \leq \epsilon&&\quad (\text{reliability})\label{eq:reliability_constraint}\\
	&I(S;W) \leq n\epsilon&&\quad(\text{secrecy})\label{eq:secrecyleakage_constraint}\\
	&H(S)\geq n(R_s-\epsilon)&&\quad(\text{key uniformity}) \label{eq:uniformity_constraint}\\
	&\log\big|\mathcal{W}\big| \leq n(R_w+\epsilon)&&\quad(\text{storage})\label{eq:storage_constraint}\\
	&I(X^n;W) \leq n(R_\ell+\epsilon)&&\quad(\text{privacy})\label{eq:leakage_constraint}.
	\end{align}
	The \emph{key-leakage-storage} region $\mathcal{R}_{\text{gs}}$ for the GS model is the closure of the set of achievable tuples.\hfill $\lozenge$
\end{definition}

Suppose the transform-coding algorithm in \cite{OurEntropy} is applied to any PUF circuits with continuous-valued outputs to obtain $X^n$ that is almost i.i.d. according to a uniform Bernoulli random variable, i.e., $X^n\sim \text{Bern}^n(\frac{1}{2})$, and the channel $P_{Y|X}$ is a $\text{BSC}(p_A)$ for $p_A\in[0, 0.5]$. Define the binary entropy function $H_b(q)= - q\log_2 q - (1-q)\log_2 (1-q)$, and the star operation $q*p = (1-2p)q+p$ with its \emph{inverse} $q=(q*p -p)/(1-2p)$.

\begin{corollary}[\hspace{1sp}\cite{IgnaTrans}] \label{cor:binaryregion}
The key-leakage-storage region $\mathcal{R}_{\text{gs,bin}}$ of the GS model for $X^n\sim \text{Bern}^n(\frac{1}{2})$ and $P_{Y|X}\sim \text{BSC}(p_A)$ is the union over all $q\in[0,0.5]$ of the bounds 
\begin{align}
&0\leq R_s\leq 1- H_b(q*p_A)\label{eq:keybinrate}\\
&R_\ell\geq H_b(q*p_A)- H_b(q)\\\
&R_w\geq H_b(q*p_A)- H_b(q)\label{eq:BSCRegionGS}.
\end{align}
\end{corollary}

The rate tuples on the boundary of the region $\mathcal{R}_{\text{gs,bin}}$ are uniquely defined by the key vs. storage rate ratio $R_s/R_w$. We therefore use this ratio as the metric to compare our nested PSCs with previously-proposed nested PCs and other channel codes. A larger key vs. storage rate ratio suggests that the code construction is closer to an achievable point that is on the boundary of $\mathcal{R}_{\text{gs,bin}}$, which is an optimal tuple.



\section{Design of Nested PSCs Construction}\label{sec:nestedPSCforPUFs}

Polar codes convert a channel into polarized virtual bit channels by a polar transform. This transform converts an input sequence $U^n$ with frozen and unfrozen bits to a length-$n$ codeword. A polar decoder processes a noisy codeword together with the frozen bits to estimate ${U}^n$. Let $\mathcal{C}(n,\mathcal{F},G^{|\mathcal{F}|})$ denote a PC or a PSC of length $n$, where $\mathcal{F}$ is the set of indices of the frozen bits and $G^{|\mathcal{F}|}$ is the sequence of frozen bits. In the following, we extend the nested PC construction proposed in \cite{RudigerPolarExtended} for the WZ problem.
\subsection{Polar Subcodes and Randomized Construction}\label{subsec:randomizedconstruction}
PSCs are a generalization of PCs, allowing some frozen symbols to be equal to linear combinations of other symbols \cite{PolarSubcodesFirst}. Such symbols are referred as dynamically-frozen symbols (DFSs). An $(n=2^m,k)$ PSC is defined by an $(n-k)\times n$ constraint matrix $\mathbb V$ such that the last non-zero elements of its rows are located in distinct columns $j_i\in\{0,\dots,n-1\}$ for $0\leq i<n-k$. The codewords of the polar subcode are obtained as $c^{n-1}=u^{n-1}\begin{pmatrix}1&0\\1&1\end{pmatrix}^{\otimes m}$, where the values $G^{|\mathcal{F}|}$ of frozen symbols are calculated as 
\begin{equation}
\label{mDynFrozen}
    u_{j_i}=\sum_{s=0}^{j_i-1}\mathbb V_{is}u_s.
\end{equation}
Decoding of PSCs can be implemented by an SC algorithm, as well as its list and sequential decoding generalizations \cite{SCLPolar,trifonov2018score}. A simple way to obtain PSCs with good performance under list or sequential decoding with small list size is to employ a randomized construction introduced in \cite{PeterRandomizedPolarSubcode}. The construction involves three types of frozen symbols:
\begin{itemize}
    \item The indices of statically-frozen symbols (SFSs), which are a special case of DFSs, are selected as integers $j_i$, for $0\leq i<n-k-t_A-t_B$, of the least reliable virtual subchannels of the polar transform, so the $i$-th row of $\mathbb V$ has $1$ in position $j_i$ and $0$, otherwise. This corresponds to constraints $u_{j_i}=0$.
    \item The indices of type-B DFSs are selected as the integers $j_i$, for $n-k-t_A-t_B\leq i<n-k-t_A$, of the least reliable virtual subchannels that are not selected as SFSs. The $i$-th row of $\mathbb V$ has $1$ in position $j_i$ and binary uniformly-random values in positions $s<j_i$. Type-B DFSs enforce the scores of incorrect paths in the Tal-Vardy decoding algorithm to decrease fast, reducing the probability of the correct path being dropped from the list.
    \item The indices of type-A DFSs  $j_i, n-k-t_A\leq j_i<n-k$, are selected as the largest integers in $\{0,1,\dots,n-1\}\setminus\{j_0,\dots,j_{n-k-t_A-1}\}$ that have the smallest weight, defined as the number of non-zero bits in a sequence's binary representation. The $i$-th row of $\mathbb V$ has $1$ in position $j_i$ and binary uniformly-random values in positions $s<j_i$. Type-A DFSs eliminate the low-weight codewords.
\end{itemize}
The number $t_A$ of type-A DFSs and the number $t_B$ of type-B DFSs should be chosen by extensive simulations. For simplicity, we use the suggested parameters for $L=32$ in \cite{kern2014new}, where $t_A=\min\{m, n-k\}$ and $t_B\!=\!\max\{0, \min\{64-t_A,n-k-t_A\}\}$. To obtain the reliabilities of the subchannels of the polar transform, we use the min-sum density evolution algorithm \cite{MinSum} over a BSC$(p)$, where the crossover probability $p$ is a design parameter to be optimized, in general, by simulations. One parameter used in the sequential decoder is the priority queue size $D$ \cite{trifonov2018score}, for which we use $D=1024$.
\subsection{Randomized Nested PSC Construction}
PCs, including PSCs, provide a simple nested code design due to the control on the subsets of codewords by changing the frozen bits. We summarise the nested code construction method proposed for PCs and then extend it to PSCs. We also provide a design procedure to design nested PSCs for key agreement with PUFs.

For the GS model with source and channel models given in Corollary~\ref{cor:binaryregion}, consider two PCs $\mathcal{C}_1(n,\mathcal{F}_1, V)$ and $\mathcal{C}(n,\mathcal{F}, \xoverline{V})$ with $\mathcal{F}=\mathcal{F}_1 \cup \mathcal{F}_w$ and $\xoverline{V}=[V, W]$, where $V$ has length $m_1$ and $W$ has length $m_2$ such that $m_1$ and $m_2$ satisfy
\begin{align}
&\frac{m_1}{n} = H_b(q)-\delta,\qquad\frac{m_1+m_2}{n} = H_b(q*p_A)+\delta\label{eq:constraintonm2and m1andm2}
\end{align}
for some distortion $q\in [0, 0.5]$ as in (\ref{eq:keybinrate})-(\ref{eq:BSCRegionGS}) and any $\delta>0$. 

We remark that (\ref{eq:constraintonm2and m1andm2}) implies a vector quantization (VQ) code $\mathcal{C}_1$ that can achieve an average per-letter distortion of at most $q$ when $n\rightarrow\infty$ since its rate is greater than the rate-distortion function $I(X;X_q)=1-H_b(q)$ at distortion $q$. (\ref{eq:constraintonm2and m1andm2}) also implies an error-correcting code (ECC) $\mathcal{C}$ that can achieve a negligible error probability for a BSC$(q*p_A)$ when $n\rightarrow\infty$ since its rate is smaller than the channel capacity $I(X_q;Y)=1-H_b(q*p_A)$. 

During enrollment, the encoder treats the uniform binary sequence $X^n$ as a noisy observation through a BSC$(q)$. Decoder of the PC $\mathcal{C}_1$ quantizes $X^n$ to a codeword $X_q^n$ of $\mathcal{C}_1$. Applying the inverse polar transform to $X_q^n$, the encoder calculates $U^n$ and its bits at indices $\mathcal{F}_w$ are stored as the helper data $W$. Furthermore, the bits at the indices $i\in \{1,2,\ldots,n\}\setminus \mathcal{F}$ are used as the secret key $S$ that has a length of $n-m_1-m_2$. 

During reconstruction, the decoder of the PC $\mathcal{C}$ observes the helper data $W$ and the binary sequence $Y^n$. The frozen bits $\xoverline{V}=[V, W]$ at indices $\mathcal{F}$ and $Y^n$ are input to the PC decoder to obtain the codeword $\widehat{X}_q^n$. Applying the inverse polar transform to $\widehat{X}_q^n$, we obtain $\widehat{U}^n$ that contains the estimate $\widehat{S}$ of the secret key at the indices $i\in \{1,2,\ldots,n\}\setminus \mathcal{F}$.

We extend the nested PC construction to nested PSCs by providing exact design parameters. We observe from simulations that VQ performance of PSCs are entirely similar to the performance of PCs, so we use a PC as the code $\mathcal{C}_1$. Let $\mathbb V_S'$ be the constraint matrix for the code $\mathcal{C}_1$, i.e., $\mathbb V_S'$ contains unit vectors with $1$s in positions $\mathcal F_1$. Then, we ensure that the low-rate PSC $\mathcal{C}$ has SFSs in indices $\mathcal F_1$. Hence, the constraint matrix $\mathbb V$ of $\mathcal{C}$ is given by 
\begin{align}
  \mathbb V=[{(\mathbb{V}_S^{'})}^T,\; {({\mathbb V}_S^{''})}^T,\;{({\mathbb V}_B)}^T,\;{({\mathbb V}_A)}^T]^T
\end{align}
where $T$ represents matrix transpose, $\mathbb V_A$ and $\mathbb V_B$ are submatrices corresponding to type-A and type-B DFSs, respectively, and $\mathbb V_S''$ corresponds to further SFSs of $\mathcal{C}$. Denote $\mathcal{F}=\mathcal{F}_A\cup\mathcal{F}_B\cup\mathcal{F}_S$ as the union of the set of indices for type-A DFSs, type-B DFSs, and all SFSs of $\mathcal{C}$.

\subsection{Proposed Design Procedure}\label{subsec:designprocedure}
We propose the following steps to design nested PSCs for source and channel models given in Corollary~\ref{cor:binaryregion} with a given blocklength $n$, secret-key size $n-m_1-m_2$, and a block-error probability $P_B$. These steps provide exact design parameter choices for nested PSCs, decided according to the simulation results over a large set of design parameters.

\begin{figure}[t] 
	\centering
	\newlength\figureheight
	\newlength\figurewidth
	\setlength\figureheight{7.75cm}
	\setlength\figurewidth{18.4cm}
	\input{./PBvspc.tikz}
	\caption{Block-error probability of $\mathcal{C}$ over a BSC with crossover probabilities $\widetilde{p}$ for Codes 1 and 2 with sequential decoders  and corresponding $\widebar{p}_c$ values represented by a circle for list size $L=8$, square for $L=32$, and pentagon for $L=64$.}	\label{fig:BLERcurve}
\end{figure}
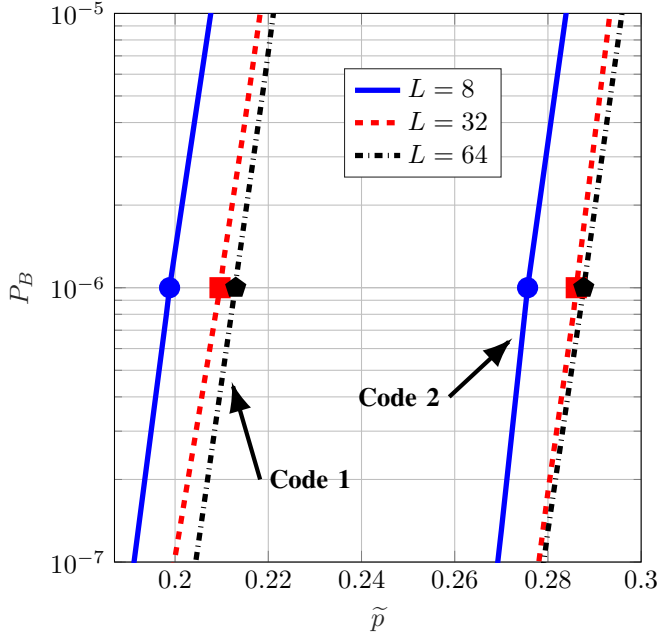 

\begin{enumerate}
	\item Apply the randomized PSC construction method given in Section~\ref{subsec:randomizedconstruction} to construct PSCs with rate $(n\!-\!m_1\!-\!m_2)/n$ for a BSC$(p)$ for a range of values in $p\in (p_A,0.5]$.\label{item:designcodesfirststep}
	\item Evaluate $P_B$ of constructed PSCs with the sequential decoder in \cite{trifonov2018score} with list size $L$ over a BSC with a range of crossover probabilities $\widetilde{p}\in (p_A,0.5]$ to obtain the crossover probability $p_c$, resulting in the target $P_B$. Assign the PSC that gives the largest $p_c$ as the low-rate PSC $\mathcal{C}$. Denote $\widebar{p}$ and $\widebar{p}_c$ as $p$ and $p_c$ values corresponding to $\mathcal{C}$, respectively. \label{item:evaluatecodessecondstep}
	\item Using the inverse of the star operation, obtain the expected target distortion $E[q]$ averaged over all $x^n\!\in\!\mathcal{X}^n$ as $E[q] =(\widebar{p}_c - p_A)/(1-2p_A)$. \label{item:evaluateEqthirdstep}
	\item Obtain the reliabilities of virtual subchannels of the polar transform by using the min-sum density evolution algorithm over a BSC$(\bar{p}_1)$, where $\widebar{p}_1=(\widebar{p}-p_A)/(1-2p_A).$\label{item:p0bar}
	\item Arrange the subchannel reliabilities obtained in Step~\ref{item:p0bar} in a descending order. Consecutively remove indices from the set $\mathcal{F}$, starting from the most reliable subchannels, until an average distortion $\widebar{q}=\frac{1}{n}\sum_{i=1}^n X_i\xor X_{q,i}$ of at most $E[q]$ is achieved, where $\xor$ denotes modulo-2 summation. Assign the remaining indices, i.e., the unremoved least reliable subchannel indices, as the frozen symbol indices of the high-rate code $\mathcal{C}_1$, denoted as $\mathcal{F}_1$.\label{item:generatehighratecodelaststep}
\end{enumerate}
Step~\ref{item:p0bar} suggests that the design parameter $\widebar{p}$ of $\mathcal{C}$ determines the design parameter $\widebar{p}_1$ for $\mathcal{C}_1$. The total number of DFSs of $\mathcal{C}$ is $t_A+t_B$, as defined in Section~\ref{subsec:randomizedconstruction}. Therefore, if the difference between the rate of $\mathcal{C}_1$ and of $\mathcal{C}$, i.e., $\Delta R\triangleq H_b(q*p_A)-H_b(q)$, is larger than $(t_A+t_B)/n$, then $\mathcal{C}_1$ is a PC because DFSs are the most reliable frozen symbols. The difference $n\Delta R$ is larger than $t_A+t_B$ for the SRAM PUF parameters we consider in the next section as $\Delta R$ increases with increasing $p_A$.

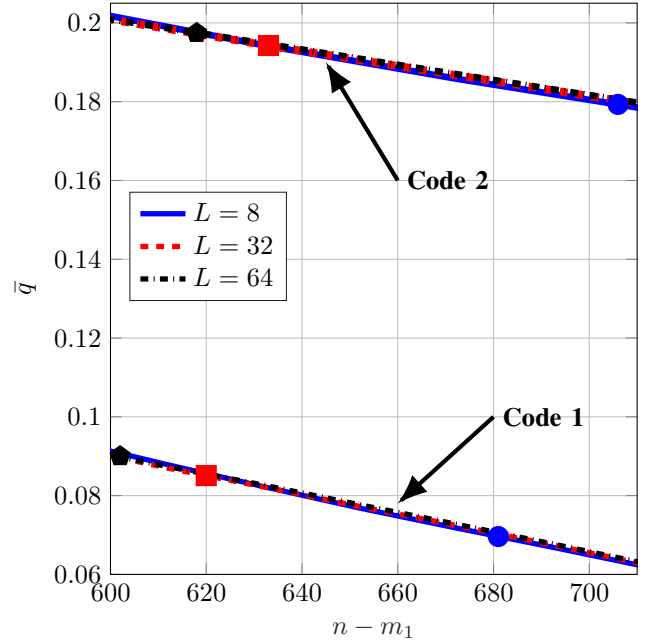
\begin{figure}[t] 
	\centering
	\setlength\figureheight{7.75cm}
	\setlength\figurewidth{18.4cm}
	\input{./Eqvsn_m0.tikz}
	\caption{Average distortion $\widebar{q}$ with respect to $n-m_1$ for Codes 1 and 2 with sequential decoders and corresponding $E[q]$ values represented by a circle for list size $L=8$, square for $L=32$, and pentagon for $L=64$.}	\label{fig:Eqcurve}
\end{figure} 

\begin{remark}
This randomized nested PSC construction provides additional degree of freedom such that the same code can be used for different  $P_B$ values or for different crossover values $p_A$ by adapting the expected distortion level.
\end{remark}

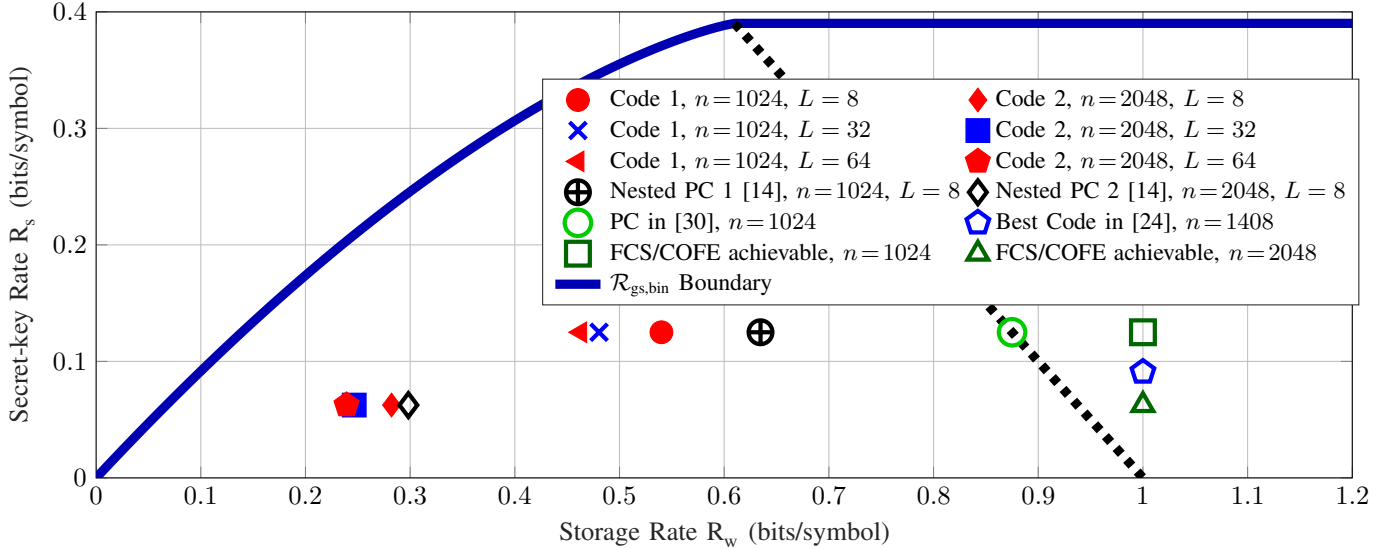
\begin{figure*}[t] 
	\centering
	\setlength\figureheight{7.75cm}
	\setlength\figurewidth{18.4cm}
	\input{./RateComparisons.tikz}
	\caption{Storage-key rates for key generation with crossover probability $p_A=0.15$. The block-error probability satisfies $P_B \leq 10^{-6}$ and the key length is 128 bits for all code points. The dashed line represents $R_w+R_s = H(X) = 1$ bit/symbol. All codes with $R_w=1$ bit/symbol are ECCs. The PC on the dashed line is a syndrome-coding construction.}
	\label{fig:codecomparisons}
\end{figure*} 

\section{Proposed Nested PSCs for PUFs}\label{sec:SRAMPUFperformance}
Consider the scenario where we generate a secret key $S$ with length $n\!-\!m_1\!-\!m_2 \!=\!128$ bits to use it in the advanced encryption standard (AES). Suppose intellectual property in a field-programmable gate array (FPGA) with an SRAM PUF should be protected so that the target block-error probability $P_B$ is $10^{-6}$ \cite{FPGAPUF}. SRAM PUF measurement channels $P_{Y|X}$ are modeled as a BSC$(p_A\!=\!0.15)$ \cite{maes2009soft}. We apply the design procedure proposed in Section~\ref{subsec:designprocedure} for these parameters to design Codes 1 and 2 with sequential decoders for list sizes $L=[8,32,64]$.

\subsubsection*{Code 1} Consider nested PSCs with blocklength $n=1024$. First, design the code $\mathcal{C}$ of rate $128/1024$ by applying Steps~\ref{item:designcodesfirststep} and \ref{item:evaluatecodessecondstep} in Section~\ref{subsec:designprocedure} for $L=8$ and obtain $\widebar{p}$. Fig.~\ref{fig:BLERcurve} depicts the $\widetilde{p}$ vs. $P_B$ curves for code $\mathcal{C}$ with sequential decoders with list sizes $L=[8,32,64]$. We observe $P_B\!=\!10^{-6}$ in Fig.~\ref{fig:BLERcurve} at crossover probabilities of $\widebar{p}_c=[0.1988,0.2096,0.2130]$ such that we obtain $E[q]=[0.0697,0.0852,0.0900]$ by Step~\ref{item:evaluateEqthirdstep} for $L = [8, 32, 64]$, respectively, where we apply $\widebar{p}$ found for $L=8$ to all list sizes for simplicity. Applying Step~\ref{item:p0bar}, we obtain the design parameter for the code $\mathcal{C}_1$ and evaluate the average distortion $\widebar{q}$ by applying Step~\ref{item:generatehighratecodelaststep}. Fig.~\ref{fig:Eqcurve} depicts the $n-m_1$ vs. $\widebar{q}$ curves obtained by applying Step~\ref{item:generatehighratecodelaststep}. Code 1 achieves $\widebar{q}=E[q]$ in Fig.~\ref{fig:Eqcurve} at $m_2=[553,492,474]$ bits of helper data, sufficing to reconstruct a $128$-bit secret key with $P_B=10^{-6}$ for $L = [8, 32, 64]$, respectively. 

\subsubsection*{Code 2} Consider nested PSCs with the same parameters as in Code 1, except $n=2048$. Fig.~\ref{fig:BLERcurve} shows that crossover probabilities of $\widebar{p}_c=[0.2756, 0.2861, 0.2883]$ satisfy $P_B=10^{-6}$, so the expected target distortions are $E[q]=[0.1795,0.1944,0.1975]$ for $L = [8, 32, 64]$, respectively. Code 2 achieves $\widebar{q}=E[q]$ in Fig.~\ref{fig:Eqcurve} at $m_2=[578,505,490]$ bits, which should be stored as helper data to generate a key size of $128$ bits with $P_B=10^{-6}$ for $L = [8, 32, 64]$, respectively.


\subsection{Rate Region Performance}
We evaluate the key-leakage-storage region $\mathcal{R}_{\text{gs,bin}}$ for $p_A=0.15$ and plot its storage-key $(R_w,R_s)$ projection in Fig.~\ref{fig:codecomparisons}. Furthermore, we plot in Fig.~\ref{fig:codecomparisons} the tuples achieved by Codes 1 and 2, previous nested PCs given in \cite{bizimWZ}, the syndrome-coding construction proposed in \cite{IgnaPolar}, and by the classic constructions that are code-offset fuzzy extractors (COFE) \cite{Dodis2008fuzzy} and the fuzzy commitment scheme (FCS) \cite{FuzzyCommitment}.

We observe from Fig.~\ref{fig:codecomparisons} that Code 1 with $L=8$ achieves a key vs. storage rate ratio of $0.2315$, improving on the nested PC 1 ratio of $0.1969$ achieved in \cite{bizimWZ} with the same list size. This result illustrates that nested PSCs achieve the best key vs. storage ratio in the literature for the same list size. Furthermore, increasing the list size of Code 1 to $L=32$ allows to achieve a ratio of $0.2602$, which is a substantial gain as compared to $L=8$ case. Further increase in the list size does not improve the achieved ratio significantly, where Code 1 with $L=64$ achieves $0.2698$. This result might be due to the choice of the numbers $t_A$ and $t_B$ of type-A and type-B DFSs adapted to $L=32$, so one might improve the performance of larger list sizes by choosing different $t_A$ and $t_B$. Similarly, Code 2 with $L=8$ achieves a $R_s/R_w$ ratio of $0.2215$, better than $0.2095$ achieved by the nested PC 2 proposed in \cite{bizimWZ}. The ratio increases to $0.2535$ and $0.2612$ by increasing the list size to $L=32$ and $L=64$, respectively. Thus, the largest $R_s/R_w$ ratio in the literature for SRAM PUFs is achieved by Code 1, for which $n=1024$ bits, with $L=64$. Its performance might be improved by optimising $t_A$ and $t_B$.

The decoding complexity of the sequential decoding algorithm in \cite{trifonov2018score} depends on the quality of the measurement channel, which depends on $p_A$ for our model. It is upper bounded by the complexity $O(Ln \log_2 n)$ of the SCL decoder, where $L$ is the maximal number of times the decoder is allowed to visit each phase (equivalent to the list size in the Tal-Vardy SCL decoding algorithm \cite{SCLPolar} used for nested PCs), but it converges to $O(n\log_2 n)$ fast with a channel bit error rate approaching 0, e.g., when $p_A\!\rightarrow\! 0$ for our model as in \cite{OurICASSP2020}. We list the average number of summation (SumCount) and comparison (CompCount) operations done with the sequential decoder of \cite{trifonov2018score} for all designed codes in Table~\ref{tab:decodingcomplexities}, where the low-rate PSCs are averaged over $10^8$ iterations and the high-rate PCs are averaged over  $20000$ iterations. Table~\ref{tab:decodingcomplexities} shows that increasing the list size $L$ or the blocklength $n$ significantly increases the decoding complexity for high-rate PCs. However, for the low-rate PSCs, increasing the list size $L$ does not increase the decoding complexity significantly; whereas increasing the blocklength $n$ has a similar effect on the decoding complexity as for high-rate PCs. Furthermore, low-rate PSCs have significantly lower decoding complexities than of high-rate PCs with the same $L$ and $n$.

\begin{table*}
	\caption{The average number of summation and comparison operations done with the sequential decoder of \cite{trifonov2018score}.}
	\centering
	\begin{tabular}{ c  C{1.7cm}  |C{1.7cm} C{1.7cm} C{1.7cm} | C{1.7cm} C{1.7cm} C{1.7cm} | }
		\cline{3-8}
		& & &  \emph{Code 1} &  &  & \emph{Code 2}&  \\
		\cline{3-8}
		& & \multicolumn{1}{c|}{$\mathbf{L=8}$} & \multicolumn{1}{c|}{$\mathbf{L=32}$} &\multicolumn{1}{c|}{$\mathbf{L=64}$} & \multicolumn{1}{c|}{$\mathbf{L=8}$} &  \multicolumn{1}{c|}{$\mathbf{L=32}$} & \multicolumn{1}{c|}{ $\mathbf{L=64}$}  \\
		& & \multicolumn{1}{c|}{$\widebar{p}_c=0.1988$} & \multicolumn{1}{c|}{$\widebar{p}_c=0.2096$} & \multicolumn{1}{c|}{$\widebar{p}_c=0.2130$} & \multicolumn{1}{c|}{$\widebar{p}_c=0.2756$} &  \multicolumn{1}{c|}{$\widebar{p}_c= 0.2861$} & \multicolumn{1}{c|}{ $\widebar{p}_c=0.2883$}  \\
		\cline{1-8}
		\multicolumn{1}{|c|}{\emph{High-rate}} & \textbf{SumCount} &  \multicolumn{1}{c|}{18596.4}& \multicolumn{1}{c|}{39404.0} & \multicolumn{1}{c|}{51431.1} & \multicolumn{1}{c|}{40893.7} & \multicolumn{1}{c|}{89803.8} & \multicolumn{1}{c|}{108000.0}\\
		\cline{2-8}
		\multicolumn{1}{|c|}{\emph{PC}} &  \textbf{CompCount} & \multicolumn{1}{c|}{14161.5} & \multicolumn{1}{c|}{29576.1}& \multicolumn{1}{c|}{38358.0} & \multicolumn{1}{c|}{33904.0} & \multicolumn{1}{c|}{73957.6} & \multicolumn{1}{c|}{88358.1}\\
		\cline{1-8}
		\multicolumn{1}{|c|}{\emph{Low-rate}} & \textbf{SumCount} &\multicolumn{1}{c|}{6512.3} &  \multicolumn{1}{c|}{6612.5} & \multicolumn{1}{c|}{6681.3} & \multicolumn{1}{c|}{13875.3} & \multicolumn{1}{c|}{14185.9} & \multicolumn{1}{c|}{14310.1}\\
		\cline{2-8}
		\multicolumn{1}{|c|}{\emph{PSC}} & \textbf{CompCount} & \multicolumn{1}{c|}{6176.4} &  \multicolumn{1}{c|}{6258.3} & \multicolumn{1}{c|}{6315.3} & \multicolumn{1}{c|}{13422.4} & \multicolumn{1}{c|}{13685.8} & \multicolumn{1}{c|}{13791.6}\\
		\cline{1-8}
	\end{tabular}\label{tab:decodingcomplexities}
\end{table*}

\section{Conclusion}\label{sec:conclusion}
We proposed a randomized nested polar subcode construction, which can be useful also for other information-theoretic problems. We proposed a design procedure to use a polar subcode as an error-correcting code and a polar code as a vector quantizer such that the codes are nested. Nested polar subcodes are designed for the source and channel models used for SRAM PUFs to illustrate significant gains in terms of the key vs. storage rate ratio as compared to previous code designs including nested polar codes. In future work, we will propose other code constructions that can perform close to the finite-length bounds one can straightforwardly calculate by combining the separate finite-length bounds for error correction and for vector quantization, which are valid also for nested code constructions considered.

\section*{Acknowledgment}
O. G\"unl\"u, M. Kim, and R. F. Schaefer were supported by the German Federal Ministry of Education and Research (BMBF) within the national initiative for ``Post Shannon Communication (NewCom)'' under the Grant 16KIS1004. V. Sidorenko is on leave from the Institute for Information Transmission Problems, Russian Academy of Science. His work was supported by the European Research Council (ERC) under the European Union’s Horizon 2020 research and innovation programme (Grant Agreement No: 801434) and by the Chair of Communications Engineering at TU Munich. O. G\"unl\"u thanks Onurcan Iscan for his insightful comments.

\bibliographystyle{IEEEtran}
\bibliography{IEEEabrv,references}

\end{document}

%% file: PBvspc.tikz
%
%
\begin{tikzpicture}

\begin{axis}[%
width=7.00156cm,
height=7.3cm,
at={(0cm,0cm)},
scale only axis,
xmin=0.187,
xmax=0.3,
xlabel style={font=\color{white!15!black}},
xlabel={$\widetilde{p}$},
ymode=log,
ymin=1e-07,
ymax=1e-05,
yminorticks=true,
ylabel style={font=\color{white!15!black}},
ylabel={$P_B$},
axis background/.style={fill=white},
xmajorgrids,
ymajorgrids,
yminorgrids,
legend style={at={(0.73506,0.90)}, legend cell align=left, align=left, draw=white!15!black}
]
\addplot [color=blue, line width=2.0pt]
  table[row sep=crcr]{%
0.19	7e-08\\
0.1981	7.9e-07\\
0.19882	1e-06\\
0.215	6.75e-05\\
0.23	0.00163953\\
0.23	0.00163953\\
0.25	0.0410341\\
0.295976	0.917431\\
};
\addlegendentry{$L=8$}

\addplot [color=red, dashed, line width=2.0pt]
  table[row sep=crcr]{%
0.19	1e-08\\
0.20964	1e-06\\
0.209783	1.05e-06\\
0.21005	1.12e-06\\
0.210423	1.26e-06\\
0.215	4.05e-06\\
0.23	0.000261\\
0.25	0.0124\\
0.278817	0.413\\
};
\addlegendentry{$L=32$}

\addplot [color=black, dashdotted, line width=2.0pt]
  table[row sep=crcr]{%
0.199733	2e-08\\
0.2	3e-08\\
0.212626	9.3e-07\\
0.2128341	1e-06\\
0.219284	5.85e-06\\
0.226888	5.48e-05\\
0.23	0.000127\\
0.231918	0.000212302\\
0.235197	0.000411093\\
0.25	0.00825\\
0.273205	0.201\\
};
\addlegendentry{$L=64$}

\addplot [color=green, line width=2.0pt, draw=none, mark size=3.0pt, mark=*, mark options={solid, blue}]
table[row sep=crcr]{%
	0.19882	1e-06\\
};

\addplot [color=black, line width=2.0pt, draw=none, mark size=3.0pt, mark=square*, mark options={solid, red}]
table[row sep=crcr]{%
	0.20964	1e-06\\
};

\addplot [color=black, line width=2.0pt, draw=none, mark size=3.0pt, mark=pentagon*, mark options={solid, black}]
  table[row sep=crcr]{%
0.213017 1e-06\\
};

\node (c1leg)[right, align=left, text=black]
at (axis cs:0.21825,0.0000002) {\normalsize \textbf{Code 1}};

\draw[decoration={markings,mark=at position 1 with {\arrow[scale=1.5]{latex}}},
postaction={decorate}, ultra thick, shorten >=1.8pt] (c1leg.west) -- ($(c1leg.west)+(-6.1,0.82)$);

\node (c2leg)[right, align=left, text=black]
at (axis cs:0.23725,0.0000004) {\normalsize \textbf{Code 2}};

\draw[decoration={markings,mark=at position 1 with {\arrow[scale=1.5]{latex}}},
postaction={decorate}, ultra thick, shorten >=1.8pt] (c2leg.east) -- ($(c2leg.east)+(13.5,0.49)$);

\addplot [color=blue, solid, line width=2.0pt]
  table[row sep=crcr]{%
0.266561	4e-08\\
0.269763	1.2e-07\\
0.27	1.3e-07\\
0.275634	1e-06\\
0.289249	4.44e-05\\
0.294622	0.000180059\\
};

\addplot [color=black, line width=2.0pt, draw=none, mark size=3.0pt, mark=*, mark options={solid, blue}]
  table[row sep=crcr]{%
0.275634	1e-06\\
};

\addplot [color=red, line width=2.0pt, dashed,]
  table[row sep=crcr]{%
0.275	4e-08\\
0.28	1.8e-07\\
0.2855	8.1e-07\\
0.28608	1e-06\\
0.298	4.45e-05\\
0.305	0.000320003\\
};

\addplot [color=black, line width=2.0pt, draw=none, mark size=3.0pt, mark=square*, mark options={solid, red}]
  table[row sep=crcr]{%
0.28608	1e-06\\
};

\addplot [color=black, dashdotted, line width=2.0pt]
  table[row sep=crcr]{%
0.275	3e-08\\
0.28	1.3e-07\\
0.287688	1e-06\\
0.298	1.81e-05\\
0.305	0.000164634\\
};

\addplot [color=black, line width=2.0pt, draw=none, mark size=3.0pt, mark=pentagon*, mark options={solid, black}]
  table[row sep=crcr]{%
0.287688	1e-06\\
};

\end{axis}

\end{tikzpicture}%

%% file: Eqvsn_m0.tikz
%
%
\begin{tikzpicture}

\begin{axis}[%
width=7.00156cm,
height=7.6cm,
at={(0cm,0cm)},
scale only axis,
xmin=600,
xmax=710,
xlabel style={font=\color{white!15!black}},
xlabel={$n-m_1$},
ymin=0.06,
ymax=0.205,
ylabel={$\widebar{q}$},
axis background/.style={fill=white},
xmajorgrids,
ymajorgrids,
yticklabel style={
	/pgf/number format/fixed,
	/pgf/number format/precision=5
},
legend style={at={(0.33506,0.67)},legend cell align=left, align=left, draw=white!15!black}
]
\addplot [color=blue, line width=2pt]
  table[row sep=crcr]{%
192	0.261584\\
212	0.251759\\
232	0.240999\\
252	0.229714\\
272	0.220456\\
292	0.212043\\
312	0.201759\\
332	0.191799\\
352	0.182327\\
372	0.173117\\
392	0.164275\\
412	0.156074\\
432	0.148397\\
452	0.140911\\
472	0.133743\\
492	0.126518\\
512	0.119477\\
532	0.112805\\
552	0.106075\\
572	0.0999517\\
592	0.0936606\\
612	0.0878241\\
632	0.0822786\\
652	0.0768424\\
672	0.0718286\\
692	0.0669811\\
712	0.0619343\\
732	0.0572172\\
752	0.053197\\
772	0.0494953\\
792	0.0461039\\
812	0.0427964\\
832	0.0410437\\
852	0.0432334\\
872	0.0462392\\
892	0.0461201\\
912	0.0410436\\
932	0.0390978\\
952	0.0318816\\
972	0.0295422\\
992	0.0192166\\
1012	0.012182\\
};
\addlegendentry{$L=8$}

\addplot [color=red, dashed, line width=2.0pt]
  table[row sep=crcr]{%
192	0.259112\\
212	0.247998\\
232	0.236948\\
252	0.225904\\
272	0.216543\\
292	0.208004\\
312	0.19795\\
332	0.188174\\
352	0.178832\\
372	0.169884\\
392	0.161198\\
412	0.153086\\
432	0.145372\\
452	0.137979\\
472	0.130825\\
492	0.123673\\
512	0.116803\\
532	0.110352\\
552	0.103803\\
572	0.0979455\\
592	0.0922074\\
612	0.0869706\\
632	0.0822044\\
652	0.0772687\\
672	0.0724806\\
692	0.0675554\\
712	0.062466\\
732	0.0577701\\
752	0.0536362\\
772	0.0498051\\
792	0.0462606\\
812	0.0427542\\
832	0.0410341\\
852	0.0427186\\
872	0.0457515\\
892	0.0432514\\
912	0.0397549\\
932	0.0374595\\
952	0.0299286\\
972	0.0261767\\
992	0.0167874\\
1012	0.00758145\\
};
\addlegendentry{$L=32$}

\addplot [color=black, dashdotted, line width=2pt]
  table[row sep=crcr]{%
192	0.258402\\
212	0.246848\\
232	0.235592\\
252	0.224693\\
272	0.215274\\
292	0.206691\\
312	0.196716\\
332	0.187002\\
352	0.177663\\
372	0.168856\\
392	0.160222\\
412	0.15211\\
432	0.144426\\
452	0.137035\\
472	0.129885\\
492	0.1228\\
512	0.115997\\
532	0.109631\\
552	0.103232\\
572	0.0976149\\
592	0.0923154\\
612	0.0873649\\
632	0.08267\\
652	0.0777006\\
672	0.0728048\\
692	0.0678282\\
712	0.0627313\\
732	0.058025\\
752	0.0538592\\
772	0.0500137\\
792	0.0464658\\
812	0.0429052\\
832	0.0411416\\
852	0.0426312\\
872	0.0456097\\
892	0.0430815\\
912	0.0394419\\
932	0.0368986\\
952	0.0291464\\
972	0.0244415\\
992	0.0162409\\
1012	0.00664297\\
};
\addlegendentry{$L=64$}

\node (c1leg)[right, align=left, text=black]
at (axis cs:680,0.1) {\normalsize \textbf{Code 1}};

\draw[decoration={markings,mark=at position 1 with {\arrow[scale=1.5]{latex}}},
postaction={decorate}, ultra thick, shorten >=1.8pt] (c1leg.west) -- ($(c1leg.west)+(-2,-22)$);

\node (c2leg)[right, align=left, text=black]
at (axis cs:660,0.16) {\normalsize \textbf{Code 2}};

\draw[decoration={markings,mark=at position 1 with {\arrow[scale=1.5]{latex}}},
postaction={decorate}, ultra thick, shorten >=1.8pt] (c2leg.west) -- ($(c2leg.west)+(-1.5,30)$);

\addplot [color=blue, line width=2.0pt, draw=none, mark size=3.0pt, mark=*, mark options={solid, blue}]
table[row sep=crcr]{%
	681	0.0696105\\
};

\addplot [color=red, line width=2.0pt, draw=none, mark size=3.0pt, mark=square*, mark options={solid, red}]
table[row sep=crcr]{%
	620	0.0850862\\
};
\addplot [color=black, line width=2.0pt, draw=none, mark size=3.0pt, mark=pentagon*, mark options={solid, black}]
  table[row sep=crcr]{%
602	0.090024\\
};

\addplot [color=blue, line width=2.5pt]
  table[row sep=crcr]{%
192	0.330532\\
232	0.314479\\
272	0.299768\\
312	0.284885\\
352	0.270622\\
392	0.257725\\
432	0.245766\\
472	0.234517\\
512	0.223728\\
552	0.213429\\
592	0.203615\\
632	0.194333\\
672	0.185874\\
712	0.178129\\
752	0.17126\\
792	0.165873\\
832	0.160807\\
872	0.156417\\
912	0.151462\\
952	0.148258\\
992	0.145612\\
1032	0.142916\\
1072	0.139652\\
1112	0.139061\\
1152	0.139066\\
1192	0.139408\\
1232	0.139342\\
1272	0.140013\\
1312	0.141259\\
1352	0.14212\\
1392	0.142989\\
1432	0.140383\\
1472	0.135106\\
1512	0.1283\\
1552	0.129722\\
1592	0.126258\\
1632	0.125806\\
1672	0.10295\\
1712	0.104625\\
1752	0.0814396\\
1792	0.0804153\\
1832	0.0817627\\
1872	0.0794865\\
1912	0.0544597\\
1952	0.0538593\\
1992	0.0267979\\
2032	0.0156875\\
};

\addplot [color=blue, line width=2.0pt, draw=none, mark size=3.0pt, mark=*, mark options={solid, blue}]
  table[row sep=crcr]{%
706	0.179309\\
};

\addplot [color=red, dashed, line width=3.0pt]
  table[row sep=crcr]{%
192	0.328526\\
232	0.311678\\
272	0.296635\\
312	0.2821\\
352	0.268057\\
392	0.255265\\
432	0.243319\\
472	0.2321\\
512	0.221488\\
552	0.211429\\
592	0.20248\\
632	0.19442\\
672	0.186752\\
712	0.179154\\
752	0.172154\\
792	0.166403\\
832	0.160962\\
872	0.156366\\
912	0.151072\\
952	0.147632\\
992	0.144437\\
1032	0.141485\\
1072	0.137987\\
1112	0.137469\\
1152	0.137744\\
1192	0.137611\\
1232	0.137618\\
1272	0.138265\\
1312	0.138852\\
1352	0.139541\\
1392	0.140278\\
1432	0.137867\\
1472	0.131999\\
1512	0.125866\\
1552	0.127061\\
1592	0.122785\\
1632	0.121056\\
1672	0.100563\\
1712	0.100749\\
1752	0.079167\\
1792	0.077096\\
1832	0.0790261\\
1872	0.0767882\\
1912	0.052335\\
1952	0.0488681\\
1992	0.0249716\\
2032	0.0136821\\
};

\addplot [color=red, line width=2.0pt, draw=none, mark size=3.0pt, mark=square*, mark options={solid, red}]
  table[row sep=crcr]{%
633	0.194253\\
};

\addplot [color=black, dashdotted, line width=2.5pt]
  table[row sep=crcr]{%
192	0.327925\\
232	0.310773\\
272	0.295638\\
312	0.281187\\
352	0.267278\\
392	0.254524\\
432	0.242565\\
472	0.231376\\
512	0.220788\\
552	0.210925\\
592	0.202465\\
632	0.194795\\
672	0.18704\\
712	0.179391\\
752	0.172315\\
792	0.166406\\
832	0.160898\\
872	0.156267\\
912	0.150867\\
952	0.147384\\
992	0.143975\\
1032	0.140938\\
1072	0.137375\\
1112	0.136854\\
1152	0.137109\\
1192	0.137017\\
1232	0.137026\\
1272	0.137624\\
1312	0.137963\\
1352	0.138726\\
1392	0.139427\\
1432	0.136914\\
1472	0.13098\\
1512	0.124605\\
1552	0.125657\\
1592	0.121388\\
1632	0.118519\\
1672	0.0995053\\
1712	0.098707\\
1752	0.0779131\\
1792	0.074032\\
1832	0.0792278\\
1872	0.0754468\\
1912	0.0517298\\
1952	0.0474824\\
1992	0.024318\\
2032	0.0132759\\
};

\addplot [color=black, line width=2.0pt, draw=none, mark size=3.0pt, mark=pentagon*, mark options={solid, black}]
  table[row sep=crcr]{%
618	0.197458\\
};

\end{axis}

\end{tikzpicture}%

%% file: RateComparisons.tikz
%
%
\definecolor{mycolor1}{rgb}{0.00000,1.00000,1.00000}%
\definecolor{mycolor2}{rgb}{0.50000,0.50000,0.500000}%
\begin{tikzpicture}

\begin{axis}[%
width=16.7cm,
height=6.2cm,
at={(0cm,0cm)},
scale only axis,
xmin=0,
xmax=1.2,
xlabel style={font=\color{white!15!black}},
xlabel={$\text{Storage Rate R}_\text{w}$ (bits/symbol)},
ymin=0,
ymax=0.4,
ylabel style={font=\color{white!15!black}},
ylabel={$\text{Secret-key Rate R}_\text{s}$ (bits/symbol)},
axis background/.style={fill=white},
xmajorgrids,
ymajorgrids,
legend style={at={(0.35506,0.3701)}, anchor=south west, legend cell align=left, legend columns=2, align=left, draw=white!15!black}
]

\addplot [only marks, color=red, mark size=4.3pt, mark=otimes*, mark options={rotate=45,solid,red}]
  table[row sep=crcr]{%
0.54003 0.125\\
};
\addlegendentry{\small Code 1, $n\!=\!1024$, $L=8$}

\addplot [only marks, color=red, mark size=4.5pt, mark=diamond*, mark options={solid,red, solid}]
  table[row sep=crcr]{%
0.282226 0.0625\\
};
\addlegendentry{\small Code 2, $n\!=\!2048$, $L=8$}

\addplot [only marks, color=blue, mark size=4.5pt, mark=+, mark options={rotate=45,solid,blue,ultra thick}]
  table[row sep=crcr]{%
0.48046 0.125\\
};
\addlegendentry{\small Code 1, $n\!=\!1024$, $L=32$}

\addplot [only marks, color=blue, mark size=4.5pt, mark=square*, mark options={solid,blue}]
  table[row sep=crcr]{%
0.246582 0.0625\\
};
\addlegendentry{\small Code 2, $n\!=\!2048$, $L=32$}

\addplot [only marks, color=red, mark size=4.5pt, mark=triangle*, mark options={rotate=90,solid,red}]
  table[row sep=crcr]{%
0.462890625 0.125\\
};
\addlegendentry{\small Code 1, $n\!=\!1024$, $L=64$}

\addplot [only marks, color=red, mark size=5pt, mark=pentagon*, mark options={solid,red}]
  table[row sep=crcr]{%
0.2392578 0.0625\\
};
\addlegendentry{\small Code 2, $n\!=\!2048$, $L=64$}

\addplot [only marks, color=black, mark size=4.5pt, mark=otimes, mark options={rotate=45,solid,black,ultra thick}]
  table[row sep=crcr]{%
0.6347	0.125\\
};
\addlegendentry{\small Nested PC 1 \cite{bizimWZ}, $n\!=\!1024$, $L=8$}

\addplot [only marks, color=black, mark size=4.5pt, mark=diamond, mark options={solid,fill=black, black, ultra thick}]
  table[row sep=crcr]{%
0.2983	0.0625\\
};
\addlegendentry{\small Nested PC 2 \cite{bizimWZ}, $n\!=\!2048$, $L=8$}

\addplot [only marks, color=black, mark size=5pt, mark=o, mark options={solid, black!20!green, ultra thick}]
  table[row sep=crcr]{%
0.875	0.125\\
};
\addlegendentry{\small PC in \cite{IgnaPolar}, $n\!=\!1024$}

\addplot [only marks, color=mycolor2, mark size=4.5pt, mark=pentagon, mark options={solid, blue, ultra thick}] 
table[row sep=crcr]{%
	1	0.09091\\
};
\addlegendentry{\small Best Code in \cite{maes2009soft}, $n\!=\!1408$}

\addplot [only marks,color=black!60!green, mark size=4.5pt, mark=square, mark options={solid, draw=black!60!green, ultra thick}]
  table[row sep=crcr]{%
1	0.125\\
};
\addlegendentry{\small FCS/COFE achievable, $n\!=\!1024$}

\addplot [only marks, solid, color=black!60!green, mark size=4.5pt, mark=triangle, mark options={solid, draw=black!60!green, ultra thick}]
  table[row sep=crcr]{%
1	0.0625\\
};
\addlegendentry{\small FCS/COFE achievable, $n\!=\!2048$}

\addplot [color=black!30!blue, solid, line width=3.5pt]
  table[row sep=crcr]{%
  	1.20	0.3901596952836\\
  	1.19	0.3901596952836\\
  	1.18	0.3901596952836\\
  	1.17	0.3901596952836\\
  	1.16	0.3901596952836\\
  	1.15	0.3901596952836\\
  	1.14	0.3901596952836\\
  	1.13	0.3901596952836\\
  	1.12	0.3901596952836\\
  	1.11	0.3901596952836\\
1.1	0.3901596952836\\
1.09	0.3901596952836\\
1.08	0.3901596952836\\
1.07	0.3901596952836\\
1.06	0.3901596952836\\
1.05	0.3901596952836\\
1.04	0.3901596952836\\
1.03	0.3901596952836\\
1.02	0.3901596952836\\
1.01	0.3901596952836\\
1	0.3901596952836\\
0.99	0.3901596952836\\
0.98	0.3901596952836\\
0.97	0.3901596952836\\
0.96	0.3901596952836\\
0.95	0.3901596952836\\
0.94	0.3901596952836\\
0.93	0.3901596952836\\
0.92	0.3901596952836\\
0.91	0.3901596952836\\
0.9	0.3901596952836\\
0.89	0.3901596952836\\
0.88	0.3901596952836\\
0.87	0.3901596952836\\
0.86	0.3901596952836\\
0.85	0.3901596952836\\
0.84	0.3901596952836\\
0.83	0.3901596952836\\
0.82	0.3901596952836\\
0.81	0.3901596952836\\
0.8	0.3901596952836\\
0.79	0.3901596952836\\
0.78	0.3901596952836\\
0.77	0.3901596952836\\
0.76	0.3901596952836\\
0.75	0.3901596952836\\
0.74	0.3901596952836\\
0.73	0.3901596952836\\
0.72	0.3901596952836\\
0.71	0.3901596952836\\
0.7	0.3901596952836\\
0.69	0.3901596952836\\
0.68	0.3901596952836\\
0.67	0.3901596952836\\
0.66	0.3901596952836\\
0.65	0.3901596952836\\
0.64	0.3901596952836\\
0.63	0.3901596952836\\
0.62	0.3901596952836\\
0.61	0.3901596952836\\
0.6098403047164	0.3901596952836\\
0.600181528522567	0.388410713739971\\
0.59251867318159	0.386667255482909\\
0.585606648644798	0.38492929951028\\
0.579180814657001	0.38319682502072\\
0.57311549625531	0.381469811410896\\
0.567336681382343	0.379748238272809\\
0.561795472576834	0.378032085391158\\
0.556457122500953	0.37632133274074\\
0.551295655828793	0.374615960483911\\
0.546290915136015	0.372915948968074\\
0.541426801360454	0.37122127872323\\
0.536690159692666	0.369531930459562\\
0.532070039497147	0.367847885065065\\
0.527557183312127	0.366169123603213\\
0.523143662589944	0.364495627310679\\
0.518822610996644	0.362827377595081\\
0.514588024639186	0.361164356032773\\
0.510434609454224	0.359506544366679\\
0.506357662610259	0.357853924504156\\
0.502352978943283	0.356206478514897\\
0.498416776149411	0.354564188628874\\
0.494545634256909	0.352927037234309\\
0.490736446124652	0.351295006875685\\
0.48698637656481	0.349668080251788\\
0.483292828289552	0.348046240213778\\
0.479653413314478	0.346429469763305\\
0.476065928767707	0.344817752050641\\
0.472528336287491	0.343211070372853\\
0.469038744366645	0.341609408172004\\
0.465595393135045	0.340012749033378\\
0.462196641173371	0.338421076683749\\
0.458840954030065	0.336834374989655\\
0.455526894175121	0.335252627955726\\
0.452253112172761	0.33367581972302\\
0.449018338893597	0.332103934567392\\
0.445821378617627	0.330536956897892\\
0.442661102904249	0.328974871255186\\
0.439536445125547	0.327417662310003\\
0.436446395575512	0.325865314861602\\
0.43338999708131	0.324317813836275\\
0.430366341053783	0.322775144285858\\
0.427374563923601	0.32123729138628\\
0.424413843917105	0.319704240436124\\
0.421483398132333	0.318175976855217\\
0.418582479881083	0.316652486183236\\
0.415710376267447	0.315133754078346\\
0.412866405977092	0.313619766315844\\
0.410049917254853	0.312110508786838\\
0.407260286051019	0.310605967496935\\
0.404496914319085	0.309106128564959\\
0.401759228449811	0.307610978221681\\
0.399046677828229	0.306120502808571\\
0.396358733501757	0.304634688776566\\
0.393694886948942	0.303153522684867\\
0.391054648939505	0.301676991199735\\
0.388437548477391	0.300205081093326\\
0.385843131819393	0.298737779242527\\
0.383270961562737	0.297275072627817\\
0.380720615795668	0.295816948332146\\
0.378191687305699	0.294363393539825\\
0.375683782840709	0.292914395535435\\
0.373196522418571	0.291469941702751\\
0.370729538681377	0.290030019523689\\
0.368282476290726	0.288594616577254\\
0.365854991360874	0.287163720538513\\
0.36344675092681	0.285737319177585\\
0.361057432444639	0.284315400358636\\
0.358686723321829	0.282897952038897\\
0.356334320475153	0.281484962267691\\
0.353999929914301	0.280076419185476\\
0.351683266349329	0.278672311022901\\
0.349384052820272	0.277272626099875\\
0.347102020347373	0.275877352824648\\
0.344836907600524	0.274486479692911\\
0.342588460586608	0.273099995286895\\
0.340356432353561	0.271717888274501\\
0.338140582710043	0.270340147408427\\
0.335940677959708	0.26896676152531\\
0.333756490649136	0.267597719544888\\
0.331587799328563	0.266233010469164\\
0.329434388324604	0.264872623381587\\
0.327296047524234	0.263516547446238\\
0.325172572169336	0.262164771907036\\
0.323063762661179	0.260817286086951\\
0.320969424374234	0.259474079387219\\
0.318889367478785	0.258135141286584\\
0.316823406771812	0.256800461340535\\
0.314771361515677	0.255470029180563\\
0.312733055284168	0.254143834513423\\
0.310708315815489	0.252821867120408\\
0.308696974871806	0.251504116856632\\
0.306698868104989	0.25019057365032\\
0.304713834928218	0.248881227502112\\
0.302741718393126	0.247576068484374\\
0.300782365072201	0.246275086740515\\
0.298835624946154	0.24497827248432\\
0.296901351295998	0.243685615999282\\
0.294979400599606	0.242397107637951\\
0.293069632432502	0.24111273782129\\
0.291171909372684	0.239832497038034\\
0.289286096909269	0.238556375844064\\
0.287412063354774	0.237284364861781\\
0.285549679760861	0.2360164547795\\
0.28369881983736	0.234752636350839\\
0.281859359874431	0.233492900394124\\
0.280031178667703	0.232237237791796\\
0.278214157446257	0.230985639489831\\
0.276408179803315	0.229738096497164\\
0.274613131629509	0.228494599885122\\
0.272828901048614	0.227255140786857\\
0.271055378355638	0.226019710396802\\
0.269292455957147	0.224788299970112\\
0.267540028313739	0.223560900822134\\
0.265797991884574	0.222337504327866\\
0.264066245073851	0.221118101921433\\
0.26234468817917	0.219902685095567\\
0.260633223341678	0.218691245401089\\
0.258931754497939	0.217483774446405\\
0.257240187333443	0.216280263897\\
0.255558429237689	0.215080705474947\\
0.25388638926078	0.213885090958411\\
0.252223978071457	0.212693412181173\\
0.250571107916528	0.211505661032142\\
0.248927692581621	0.210321829454892\\
0.247293647353214	0.20914190944719\\
0.24566888898189	0.207965893060535\\
0.244053335646775	0.206793772399705\\
0.24244690692109	0.205625539622303\\
0.240849523738808	0.204461186938316\\
0.239261108362338	0.203300706609673\\
0.237681584351223	0.20214409094981\\
0.236110876531794	0.200991332323242\\
0.234548910967763	0.19984242314514\\
0.232995614931693	0.198697355880905\\
0.231450916877348	0.197556123045763\\
0.22991474641285	0.196418717204346\\
0.228387034274647	0.195285130970296\\
0.226867712302246	0.194155357005856\\
0.225356713413682	0.193029388021481\\
0.223853971581703	0.191907216775441\\
0.222359421810641	0.190788836073442\\
0.220873000113953	0.189674238768234\\
0.219394643492391	0.188563417759244\\
0.217924289912796	0.187456365992191\\
0.216461878287488	0.186353076458728\\
0.215007348454228	0.185253542196066\\
0.213560641156738	0.184157756286622\\
0.212121698025755	0.183065711857657\\
0.210690461560606	0.181977402080923\\
0.209266875111283	0.180892820172317\\
0.207850882860997	0.179811959391533\\
0.206442429809208	0.178734813041722\\
0.205041461755106	0.177661374469155\\
0.203647925281517	0.176591637062888\\
0.202261767739255	0.175525594254432\\
0.200882937231859	0.174463239517428\\
0.19951138260075	0.173404566367324\\
0.198147053410747	0.172349568361054\\
0.196789899935978	0.171298239096723\\
0.195439873146137	0.170250572213297\\
0.194096924693087	0.169206561390291\\
0.192761006897815	0.168166200347463\\
0.191432072737698	0.167129482844518\\
0.190110075834097	0.1660964026808\\
0.18879497044025	0.165066953695003\\
0.187486711429475	0.164041129764877\\
0.186185254283645	0.163018924806936\\
0.184890555081958	0.162000332776175\\
0.183602570489972	0.160985347665786\\
0.182321257748904	0.159973963506876\\
0.181046574665185	0.158966174368193\\
0.179778479600266	0.157961974355849\\
0.178516931460662	0.156961357613048\\
0.17726188968823	0.155964318319823\\
0.176013314250678	0.154970850692764\\
0.174771165632291	0.153980948984759\\
0.17353540482487	0.152994607484731\\
0.17230599331889	0.152011820517385\\
0.171082893094842	0.15103258244295\\
0.169866066614792	0.150056887656928\\
0.168655476814115	0.149084730589844\\
0.167451087093425	0.148116105707002\\
0.166252861310681	0.147151007508235\\
0.165060763773472	0.146189430527672\\
0.163874759231469	0.14523136933349\\
0.162694812869047	0.144276818527685\\
0.161520890298068	0.143325772745831\\
0.160352957550818	0.142378226656854\\
0.159190981073106	0.141434174962798\\
0.158034927717501	0.140493612398601\\
0.156884764736722	0.139556533731871\\
0.155740459777167	0.138622933762658\\
0.154601980872581	0.13769280732324\\
0.153469296437855	0.136766149277903\\
0.152342375262957	0.135842954522723\\
0.151221186506995	0.134923217985357\\
0.150105699692397	0.134006934624826\\
0.148995884699211	0.133094099431311\\
0.147891711759534	0.132184707425942\\
0.146793151452042	0.131278753660595\\
0.145700174696641	0.13037623321769\\
0.144612752749221	0.129477141209986\\
0.143530857196523	0.128581472780386\\
0.142454459951104	0.127689223101739\\
0.141383533246404	0.126800387376641\\
0.140318049631914	0.125914960837251\\
0.139257981968439	0.125032938745089\\
0.138203303423455	0.124154316390856\\
0.137153987466556	0.123279089094239\\
0.136110007864994	0.122407252203733\\
0.135071338679299	0.121538801096448\\
0.134037954258998	0.120673731177938\\
0.133009829238397	0.119812037882012\\
0.131986938532465	0.118953716670558\\
0.130969257332784	0.118098763033369\\
0.129956761103579	0.117247172487967\\
0.128949425577823	0.116398940579427\\
0.127947226753424	0.115554062880209\\
0.12695014088947	0.114712534989986\\
0.125958144502556	0.113874352535478\\
0.124971214363178	0.113039511170282\\
0.123989327492184	0.11220800657471\\
0.123012461157306	0.111379834455625\\
0.122040592869747	0.11055499054628\\
0.121073700380832	0.109733470606154\\
0.12011176167872	0.108915270420799\\
0.11915475498518	0.108100385801677\\
0.118202658752422	0.107288812586011\\
0.117255451659982	0.106480546636622\\
0.116313112611675	0.105675583841787\\
0.115375620732588	0.104873920115078\\
0.114442955366138	0.10407555139522\\
0.113515096071175	0.103280473645935\\
0.112592022619145	0.102488682855802\\
0.11167371499129	0.101700175038109\\
0.110760153375914	0.100914946230706\\
0.109851318165681	0.100132992495866\\
0.108947189954971	0.0993543099201439\\
0.108047749537278	0.0985788946142315\\
0.107152977902654	0.0978067427128236\\
0.106262856235194	0.0970378503744788\\
0.105377365910569	0.0962722137814831\\
0.104496488493601	0.0955098291397151\\
0.103620205735875	0.0947506926785127\\
0.102748499573395	0.0939948006505404\\
0.10188135212428	0.0932421493316584\\
0.101018745686498	0.0924927350207921\\
0.100160662735637	0.091746554039804\\
0.099307085922717	0.0910036027333663\\
0.0984579980720334	0.0902638774688338\\
0.0976133821790415	0.0895273746361198\\
0.0967732214082729	0.0887940906475712\\
0.0959374990912875	0.088064021937846\\
0.0951061987246586	0.087337164963792\\
0.0942793039679912	0.0866135162043258\\
0.093456798641975	0.0858930721603133\\
0.0926386667264648	0.0851758293544517\\
0.0918248923585963	0.0844617843311524\\
0.0910154598309288	0.0837509336564244\\
0.0902103535896233	0.0830432739177589\\
0.0894095582326417	0.0823388017240165\\
0.0886130585079838	0.0816375137053134\\
0.0878208393119454	0.0809394065129093\\
0.087032885687409	0.0802444768190971\\
0.0862491828221583	0.0795527213170925\\
0.085469716047222	0.0788641367209255\\
0.0846944708352413	0.0781787197653325\\
0.0839234327988658	0.0774964672056493\\
0.0831565876891714	0.0768173758177051\\
0.0823939213941066	0.0761414423977177\\
0.0816354199369596	0.0754686637621897\\
0.0808810694748507	0.0747990367478055\\
0.0801308562972489	0.0741325582113288\\
0.0793847668245108	0.0734692250295015\\
0.0786427876064396	0.0728090340989445\\
0.0779049053208714	0.0721519823360568\\
0.0771711067722778	0.0714980666769185\\
0.0764413788903932	0.0708472840771923\\
0.0757157087288609	0.0701996315120283\\
0.0749940834639036	0.0695551059759656\\
0.0742764903930073	0.068913704482841\\
0.073562916933632	0.0682754240656915\\
0.0728533506219368	0.0676402617766645\\
0.0721477791115278	0.0670082146869224\\
0.0714461901722212	0.0663792798865539\\
0.0707485716888285	0.0657534544844809\\
0.070054911659957	0.0651307356083705\\
0.0693651981968286	0.0645111204045454\\
0.0686794195221175	0.0638946060378958\\
0.0679975639688036	0.0632811896917918\\
0.067319619979043	0.0626708685679974\\
0.066645576103056	0.0620636398865841\\
0.0659754209980307	0.0614595008858458\\
0.0653091434270419	0.0608584488222155\\
0.0646467322579873	0.0602604809701807\\
0.0639881764625378	0.0596655946222016\\
0.0633334651151037	0.059073787088629\\
0.0626825873918166	0.058485055697622\\
0.062035532569524	0.0578993977950693\\
0.0613922900247995	0.0573168107445078\\
0.0607528492329686	0.0567372919270448\\
0.0601171997671454	0.0561608387412786\\
0.0594853312972856	0.0555874486032224\\
0.058857233589252	0.0550171189462259\\
0.0582328965038934	0.0544498472208998\\
0.0576123099961372	0.0538856308950402\\
0.0569954641140928	0.0533244674535542\\
0.0563823489981713	0.0527663543983847\\
0.0557729548802128	0.0522112892484385\\
0.0551672720826311	0.0516592695395122\\
0.0545652910175669	0.0511102928242202\\
0.0539670021860534	0.0505643566719243\\
0.0533723961771955	0.0500214586686618\\
0.0527814636673574	0.0494815964170756\\
0.0521941954193645	0.0489447675363452\\
0.0516105822817143	0.0484109696621171\\
0.0510306151878002	0.0478802004464367\\
0.0504542851551424	0.0473524575576811\\
0.0498815832846349	0.0468277386804918\\
0.0493125007597979	0.0463060415157082\\
0.0487470288460434	0.0457873637803017\\
0.0481851588899499	0.0452717032073108\\
0.0476268823185465	0.044759057545777\\
0.0470721906386085	0.0442494245606795\\
0.046521075435961	0.043742802032873\\
0.0459735283747928	0.0432391877590239\\
0.0454295411969798	0.0427385795515496\\
0.0448891057214169	0.0422409752385544\\
0.0443522138433595	0.0417463726637711\\
0.0438188575337741	0.0412547696864984\\
0.0432890288386971	0.0407661641815422\\
0.0427627198786036	0.0402805540391553\\
0.0422399228477822	0.0397979371649793\\
0.0417206300137212	0.0393183114799858\\
0.0412048337165014	0.0388416749204188\\
0.0406925263681972	0.0383680254377373\\
0.0401837004522858	0.0378973609985591\\
0.0396783485230654	0.0374296795846041\\
0.0391764632050792	0.0369649791926386\\
0.0386780371925485	0.0365032578344215\\
0.0381830632488149	0.0360445135366469\\
0.0376915342057856	0.0355887443408932\\
0.0372034429633903	0.0351359483035678\\
0.0367187824890437	0.0346861234958539\\
0.0362375458171147	0.034239268003659\\
0.0357597260484037	0.0337953799275619\\
0.0352853163496272	0.0333544573827615\\
0.0348143099529076	0.0329164984990254\\
0.03434670015527	0.0324815014206403\\
0.0338824803181484	0.0320494643063606\\
0.0334216438668951	0.0316203853293596\\
0.032964184290298	0.0311942626771804\\
0.0325100951401053	0.0307710945516868\\
0.0320593700305548	0.0303508791690148\\
0.0316120026379103	0.0299336147595259\\
0.0311679867000045	0.0295192995677593\\
0.0307273160157867	0.029107931852385\\
0.0302899844448776	0.0286995098861578\\
0.029855985907131	0.0282940319558704\\
0.0294253143821986	0.0278914963623091\\
0.0289979639091025	0.0274919014202086\\
0.028573928585813	0.027095245458207\\
0.0281532025688317	0.0267015268188021\\
0.0277357800727807	0.0263107438583077\\
0.0273216553699973	0.0259228949468095\\
0.0269108227901327	0.025537978468124\\
0.0265032767197577	0.0251559928197544\\
0.0260990116019729	0.0247769364128497\\
0.0256980219360246	0.0244008076721629\\
0.0253003022769244	0.0240276050360098\\
0.0249058472350756	0.0236573269562284\\
0.0245146514759036	0.0232899718981388\\
0.0241267097194928	0.0229255383405023\\
0.0237420167402242	0.0225640247754839\\
0.0233605673664233	0.0222054297086116\\
0.0229823564800091	0.0218497516587384\\
0.0226073790161486	0.0214969891580038\\
0.0222356299629167	0.0211471407517965\\
0.0218671043609595	0.0208002049987162\\
0.0215017973031632	0.0204561804705373\\
0.0211397039343268	0.0201150657521716\\
0.0207808194508396	0.0197768594416328\\
0.020425139100363	0.0194415601499993\\
0.0200726581815157	0.0191091665013804\\
0.0197233720435648	0.0187796771328801\\
0.0193772760861197	0.0184530906945624\\
0.0190343657588316	0.0181294058494172\\
0.0186946365610946	0.0178086212733268\\
0.0183580840417543	0.0174907356550309\\
0.0180247037988178	0.017175747696095\\
0.0176944914791676	0.0168636561108765\\
0.0173674427782825	0.016554459626492\\
0.0170435534399573	0.016248156982787\\
0.0167228192560323	0.0159447469323012\\
0.0164052360661212	0.0156442282402394\\
0.0160907997573463	0.0153465996844397\\
0.0157795062640753	0.0150518600553429\\
0.0154713515676641	0.0147600081559617\\
0.0151663316962	0.0144710428018515\\
0.0148644427242512	0.0141849628210802\\
0.0145656807726198	0.0139017670541988\\
0.0142700420080965	0.0136214543542126\\
0.0139775226432193	0.0133440235865532\\
0.0136881189360365	0.0130694736290498\\
0.0134018271898743	0.0127978033719004\\
0.0131186437531035	0.012529011717646\\
0.0128385650189137	0.0122630975811419\\
0.0125615874250903	0.0120000598895311\\
0.0122877074537913	0.0117398975822185\\
0.0120169216313322	0.0114826096108437\\
0.0117492265279704	0.0112281949392555\\
0.0114846187576944	0.0109766525434868\\
0.0112230949780163	0.0107279814117281\\
0.0109646518897668	0.0104821805443041\\
0.010709286236893	0.0102392489536476\\
0.0104569948062606	0.00999918566427671\\
0.0102077744274577	0.00976198971276943\\
0.00996162197260175	0.00952766014774142\\
0.00971853435615055	0.0092961960298219\\
0.00947850853471632	0.00906759643163013\\
0.00924154150687961	0.00884186043775403\\
0.00900763031301033	0.00861898714472709\\
0.00877677203508931	0.00839897566100567\\
0.00854896379653236	0.00818182510694898\\
0.00832420276201828	0.00796753461479582\\
0.0081024861373199	0.00775610332864496\\
0.00788381116913617	0.00754753040443257\\
0.0076681751449289	0.00734181500991371\\
0.00745557539276054	0.00713895632464068\\
0.00724600928113628	0.00693895353994389\\
0.00703947421884799	0.00674180585891149\\
0.00683596765481953	0.0065475124963712\\
0.00663548707795858	0.00635607267887006\\
0.00643803001700571	0.00616748564465674\\
0.00624359404039088	0.00598175064366224\\
0.00605217675608938	0.0057988669374831\\
0.0058637758114819	0.00561883379936168\\
0.00567838889321681	0.00544165051417056\\
0.00549601372707387	0.00526731637839406\\
0.00531664807783272	0.00509583070011155\\
0.00514028974914094	0.00492719279898179\\
0.00496693658338732	0.00476140200622499\\
0.00479658646157655	0.0045984576646072\\
0.00462923730320397	0.00443835912842616\\
0.00446488706613768	0.00428110576349339\\
0.0043035337464985	0.00412669694712076\\
0.00414517537854475	0.00397513206810418\\
0.00398981003455823	0.00382641052671073\\
0.00383743582473328	0.00368053173466243\\
0.00368805089706792	0.00353749511512369\\
0.00354165343725787	0.00339730010268591\\
0.00339824166858982	0.00325994614335667\\
0.00325781385184265	0.00312543269454313\\
0.00312036828518614	0.00299375922504175\\
0.00298590330408277	0.00286492521502413\\
0.0028544172811944	0.00273893015602544\\
0.00272590862628741	0.00261577355093168\\
0.00260037578614258	0.00249545491396896\\
0.00247781724446738	0.00237797377069038\\
0.0023582315218087	0.0022633296579655\\
0.00224161717546822	0.00215152212396996\\
0.00212797279942134	0.00204255072817405\\
0.00201729702423714	0.00193641504133168\\
0.0019095885170004	0.00183311464547198\\
0.00180484598123565	0.00173264913388782\\
0.0017030681568353	0.00163501811112599\\
0.00160425381998641	0.00154022119297936\\
0.00150840178310219	0.00144825800647685\\
0.001415510894756	0.00135912818987416\\
0.00132558003961436	0.001272831392646\\
0.00123860813837584	0.00118936727547769\\
0.00115459414770913	0.00110873551025625\\
0.00107353706019486	0.00103093578006408\\
0.000995435904267983	0.00095596777917073\\
0.000920289744165137	0.000883831213024822\\
0.000848097679869042	0.000814525798249033\\
0.000778858847060304	0.000748051262631888\\
0.000712572417069346	0.000684407345121096\\
0.000649237596826779	0.000623593795819222\\
0.000588853628822328	0.000565610375975911\\
0.000531419791060528	0.000510456857982788\\
0.000476935397019873	0.000458133025369456\\
0.000425399795616843	0.000408638672795725\\
0.000376812371166713	0.000361973606049171\\
0.000331172543350355	0.000318137642039362\\
0.000288479767182048	0.000277130608793863\\
0.000248733532978052	0.000238952345454124\\
0.000211933366329298	0.000203602702272154\\
0.000178078828073969	0.000171081540605855\\
0.000147169514273182	0.000141388732916914\\
0.000119205056188676	0.00011452416276736\\
9.41851202626021e-05	9.04877248166791e-05\\
7.21094080974316e-05	6.92793248194823e-05\\
5.29776564414108e-05	5.08988796231735e-05\\
3.67896371719079e-05	3.5346317166618e-05\\
2.35451572835332e-05	2.26215764774773e-05\\
1.3244058878259e-05	1.27246076710996e-05\\
5.88621915487142e-06	5.65537194996413e-06\\
1.47155040330915e-06	1.41384160201596e-06\\
0	0\\
};
\addlegendentry{\small $\mathcal{R}_{\text{gs,bin}}$ Boundary}

\addplot [color=black,dashed,line width=3.5pt]
table[row sep=crcr]{%
	0.6098403047164	0.3901596952836\\
	0.6198403047164	0.3801596952836\\
	0.6298403047164	0.3701596952836\\
	0.6398403047164	0.3601596952836\\
	0.6498403047164	0.3501596952836\\
	0.6598403047164	0.3401596952836\\
	0.6698403047164	0.3301596952836\\
	0.6798403047164	0.3201596952836\\
	0.6898403047164	0.3101596952836\\
	0.6998403047164	0.3001596952836\\
	0.7098403047164	0.2901596952836\\
	0.7198403047164	0.2801596952836\\
	0.7298403047164	0.2701596952836\\
	0.7398403047164	0.2601596952836\\
	0.7498403047164	0.2501596952836\\
	0.7598403047164	0.2401596952836\\
	0.7698403047164	0.2301596952836\\
	0.7798403047164	0.2201596952836\\
	0.7898403047164	0.2101596952836\\
	0.7998403047164	0.2001596952836\\
	0.8098403047164	0.1901596952836\\
	0.8198403047164	0.1801596952836\\
	0.8298403047164	0.1701596952836\\
	0.8398403047164	0.1601596952836\\
	0.8498403047164	0.1501596952836\\
	0.8598403047164	0.1401596952836\\
	0.8698403047164	0.1301596952836\\
	0.8798403047164	0.1201596952836\\
	0.8898403047164	0.1101596952836\\
	0.8998403047164	0.1001596952836\\
	0.9098403047164	0.0901596952836\\
	0.9198403047164	0.0801596952836\\
	0.9298403047164	0.0701596952836\\
	0.9398403047164	0.0601596952836\\
	0.9498403047164	0.0501596952836\\
	0.9598403047164	0.0401596952836\\
	0.9698403047164	0.0301596952836\\
	0.9798403047164	0.0201596952836\\
	0.9898403047164	0.0101596952836\\
	0.9998403047164	0.0001596952836\\
};

\end{axis}
\end{tikzpicture}%